\documentclass[%
 reprint,standalone,
showkeys,
showpacs,preprintnumbers,
 amsmath,amssymb,
 aps,
]{revtex4-1}
\usepackage{subcaption}
\usepackage[labelformat=parens,labelsep=quad,skip=3pt]{caption}
\usepackage{graphicx}
\usepackage{hhline}
\usepackage{dcolumn}
\usepackage{bm}
\usepackage{color,soul}

\begin{document}

\preprint{Rev. Mex. Fis. E.}

\title{Revisiting viscosity from macroscopic to nanoscale regimes}

\author{G. Hamilton}
 \email{ghamil4@clemson.edu}
\author{Z. Disharoon}
 \email{zack\_disharoon@yahoo.com}
\author{H. Sanabria}
 \email{hsanabr@clemson.edu}
\affiliation{Department of Physics and Astronomy, Clemson University, Clemson, SC 29634}

\date{\today}

\begin{abstract}
The response of a fluid to deformation by shear stress is known as shear viscosity. This concept arises from a macroscopic view and was first introduced by Sir Isaac Newton. Nonetheless, a fluid is a series of moving molecules that are constrained by the shape of the container. Such a view begs the treatment of viscosity from a microscopic or molecular view, a task undertaken by both Einstein and Smoluchowski independently. Here we revisit the concept of viscosity and experimentally verify that the viscosity at a molecular level, which describes the drag force, is the same as the macroscopic shear viscosity; hence, bridging different length- and time-scales. For capturing the shear stress response of a fluid, we use classical rheometry; at a molecular level we use probe diffusion to determine the local viscosity from the translational and rotational motions. In these cases, we use Fluorescence Correlation Spectroscopy and Time Resolved Fluorescence, respectively. By increasing the osmolyte (Glucose-D) concentration, we change the viscosity and find that these methods provide a unified view of viscosity bridging the gap between the macroscopic and nanoscale regimes. Moreover, Glucose's viscosity follows a scaling factor more commonly associated to solutions of branched polymer because the probe dimensions are comparable to the dimensions of the osmolyte that exerts the drag.

\end{abstract}



\maketitle


\section{\label{sec:level1}Introduction}

The viscosity of a fluid describes the internal drag forces within the fluid as it is subjected to stress.\cite{Berg} Accurate descriptions of viscosity have a broad range of applications, from the characterization of blood flow as it relates to corollary heart disease to optimizing lubricants for mechanical systems.\cite{viscosity app blood, viscosity app lub} Isaac Newton first described fluid viscosity in his 1687 \textit{Principia}, where he stated Newton's Law of Viscosity, describing the response of a continuous, incompressible fluid to shear stress.\cite{Principia} In the 1840s, the Navier-Stokes equation was derived and used to describe the diffusion of totally conserved quantities through a continuous fluid.\cite{Navier-Stokes} In 1866, James Maxwell reported experimental results which verified his earlier calculations showing viscosity of a gas is dependent on the mean free path of its particles.\cite{Maxwell} Later, a microscopic description of diffusion grounded in Robert Brown's 1827 observations of pollen particles randomly moving was independently developed by Albert Einstein in 1905 and by Marian Smoluchowski in 1906, resulting in the Einstein-Smoluchowski relation describing the probe diffusion coefficient.\cite{Einstein, Smoluchowski} Through George Stokes' 1851 derivation of Stokes' Law, the diffusion coefficient is related back to the fluid viscosity describing the frictional drag felt by individual particles in a continuous media.\cite{Stokes}

In the broadest mathematical formalism, viscosity is given by the viscosity tensor, $\boldsymbol{\mu}$, in
\begin{equation} \label{tensor}
\boldsymbol{\tau} =2 \boldsymbol{\mu \epsilon},
\end{equation}
  relating the viscous stress tensor for a fluid, $\boldsymbol{\tau}$, and strain rate tensor, $\boldsymbol{\epsilon}$.\cite{Groot} For Newtonian fluids, $\boldsymbol{\mu}$ has three independent components: the bulk, dynamic, and rotational viscosities which describe fluid response to compressive forces, resistance to shear, and the coupling between flow and individual particle rotations, respectively.\cite{Groot} As was done historically, compatible expressions for dynamic viscosity and related quantities can be derived under two different theoretical models of fluids. The first model is a macroscopic, continuum one relating diffusion to concentration gradients. The second is a statistical, Brownian motion model of single particle diffusion at local scales. Introduction of the term viscosity from both perspectives is followed by three different approaches to experimentally determine the viscosity of a Newtonian fluid. These three experimental methods range in scales from nm to cm and from ns to seconds. The shear stress of solutions of glucose at different concentrations was studied using a rheometer. For the probing of local viscosity, translational and rotational diffusion of Rhodamine-110 were observed in glucose solution utilizing Fluorescence Correlation Spectroscopy (FCS) and time-resolved anisotropy measurements, respectively.

\subsection{\label{sec:level2}Continuous fluid}
Empirically verified by Newton, Newton's Law of Viscosity is
\begin{equation} \label{force}
\mathbf{F}/A = \eta \nabla \mathbf{v}, 
\end{equation}
where $\mathbf{F}$ is the force in contact with a liquid over a cross sectional area $A$, $\nabla \mathbf{v}$ is the velocity gradient in the fluid, and $\eta$ is defined as the shear, or dynamic, element of the viscosity tensor.\cite{Newton, Themelis viscosity form} This expression is a special case of Eq.\eqref{tensor} for shear stress applied to isotropic, incompressible Newtonian fluids, in which case $\boldsymbol{\mu}$ reduces to the dynamic viscosity, $\eta$.\cite{Landau}

The dynamics of fluids, including diffusive processes which depend on the viscosity of the fluid, can also be described using the Navier-Stokes equation. Let us begin from the Cauchy Momentum Equation, a statement of conservation of momentum for a continuum:
\begin{equation}
\frac{\partial \mathbf{p}}{\partial t} + \nabla \cdot \mathbf{J_{p}} = \mathbf{s},
\end{equation}
where $\mathbf{p}=\rho \mathbf{u}$ is the momentum density defined by the mass density ($\rho$) times velocity ($\mathbf{u}$), $t$ is time, $\mathbf{J_{p}}$ is the momentum density flux out of the volume, and $\mathbf{s}$ is a source term corresponding to stresses and forces imparting momentum on the system. Rewriting in terms of $\rho$, $\mathbf{u}$, the material derivative $\frac{D}{Dt}$, the sum of externally-caused accelerations, $\mathbf{g}$, the pressure, $p$, and the stress tensor, $\boldsymbol{\tau}$, we obtain
\begin{equation} \label{fulleqns}
 \frac{D \rho \mathbf{u}}{Dt}=\rho \mathbf{g} - \nabla p + \nabla \cdot \tau.
\end{equation}
Plugging in Eq.\eqref{tensor}, where $\boldsymbol{\epsilon}=\frac{1}{2}(\nabla \mathbf{u}+\nabla \mathbf{u}^{T})-\frac{1}{3} \nabla \cdot \mathbf{u} \delta_{ij}$, Eq.\eqref{fulleqns} we obtain
\begin{equation}
 \frac{D \rho \mathbf{u}}{Dt}=\rho \mathbf{g} - \nabla p + \nabla \cdot (2\boldsymbol{\mu}(\frac{1}{2}(\nabla \mathbf{u}+\nabla \mathbf{u}^{T})-\frac{1}{3} \nabla \cdot \mathbf{u} \delta_{ij})). 
\end{equation}
Taking the limit of an incompressible, Newtonian fluid, $\rho$ is a constant and $\nabla \cdot \mathbf{u}=0$. Further, taking $\mu$ as a constant with respect to position, $\nabla \cdot \tau$ reduces to $\mu \nabla^{2} \mathbf{u}$ and we are left with
\begin{equation}
\rho \frac{D \mathbf{u}}{Dt}=\rho \mathbf{g} - \nabla p + \mu \nabla^{2} \mathbf{u}.
\end{equation}
Finally, we take the limits in which the external and hydrostatic force contributions are negligible and there is no net fluid flow. The net result is that the material derivative reduces to the partial time derivative and the external force terms become $0$, leaving
\begin{equation} \label{NSsimp}
\rho \frac{\partial \mathbf{u}}{\partial t} =\mu \nabla^{2} \mathbf{u}.
\end{equation}
This is the linear Navier-Stokes momentum density equation for an incompressible, non-convecting fluid subjected to viscous forces.\cite{Batchelor} Eq.\eqref{NSsimp} is essentially the momentum diffusion equation and is closely related to Eq.\eqref{force}. A similar analysis can be performed beginning from the conservation of amount of fluid in terms of concentration, $C$, with no sources:

\begin{equation} \label{consmass}
\frac{\partial C}{\partial t}+\nabla \cdot \mathbf{J_{fluid}}=0.
\end{equation}
Making use of the empirically verified Fick's First Law, 
\begin{equation} \label{wut}
\mathbf{J_{fluid}}=-D\nabla C,
\end{equation}
where $D$ is defined as the translational diffusion coefficient, we obtain
\begin{equation} \label{diffeq}
\frac{\partial C}{\partial t}=D\nabla^{2}C.
\end{equation}
This is the non-convective, un-sourced diffusion equation in terms of concentration, or Fick's 
Second Law for constant $D$.\cite{Fick,Berg} $D$ describes the rate at which a substance (either a continuum  or a single particle) diffuses through medium. $D$ depends on the viscous drag force exerted by the fluid. $D$ can be explicitly related to the viscosity experienced by a probe through the Stokes-Einstein relation discussed in the next section.\cite{stokes-einstein}

\subsection{\label{sec:level2}Thermally-driven random walk}
Viscosity can also be defined by considering individual particles undergoing Brownian motion in media. In Brownian motion, each individual probe particle undergoes an effective random walk driven by thermal energy.\cite{brownian motion} This model allows a derivation of the diffusion equation from a microscopic, statistical perspective which matches the continuum case. Let us consider a particle randomly walking in space with probabilities $p$ and $q$ to travel either right or left in the $\hat{x}$ direction a distance $\delta$ during timestep $\tau$. Then the probability of finding the particle at position $x$ at time $t+\tau$ is given by
\begin{equation} \label{probs}
P(x,t+\tau)=pP(x-\delta,t)+qP(x+\delta,t).
\end{equation}
By doing a Taylor expansion to first order in $t$ and second order in $x$ in the continuous limit (small $\tau$ and $\delta$), we obtain that the first non-vanishing contribution leads to
\begin{equation} \label{probdiff}
\frac{\partial P(x,t)}{\partial t}+(p-q)\frac{\partial x}{\partial t}\frac{\partial P(x,t)}{\partial x}=D\frac{\partial^{2}P(x,t)}{\partial x^{2}},
\end{equation}
the convective diffusion equation, with translational diffusion coefficient $D=\frac{\delta^{2}}{2\tau}$.\cite{kinproc} In our case of unbiased walking, $p=q$ and this reduces to
\begin{equation} \label{probdiff2}
\frac{\partial P(x,t)}{\partial t}=D\frac{\partial^{2}P(x,t)}{\partial x^{2}}.
\end{equation}
Finally, assuming that $D$ is the same in each direction and summing over the three Cartesian coordinates, we obtain
\begin{equation}
\frac{\partial P(\mathbf{r},t)}{\partial t}=D\nabla^{2}P(\mathbf{r},t),
\end{equation}
the analog of Eq.\eqref{diffeq} for a single probe diffusing through a fluid. These equations describe the same phenomena under the substitution $P=C$, or taking particle concentration for many particles as a probability distribution. This is seen in their shared solution
\begin{equation}\label{x2}
\langle r^2 \rangle =6Dt,
\end{equation}
where $\langle r^{2} 
\rangle$ is the mean-square position for an ensemble of particles at time $t$.\cite{Berg} The ergodic principle assures us that the ensemble average $\langle ... \rangle$ is equal to the time average.\cite{hopf} An identically structured derivation can also be performed by considering an angular random walk, or the random reorientation of a probe molecule such that the probability of finding the particle in some orientation is given in an identical form to Eq.\eqref{probs}. In this case the solution
\begin{equation}
\langle \theta^2 \rangle =2D_{\theta}t,
\end{equation}
is found, where $\theta$ is the orientation angle and $D_{\theta}$ is the rotational diffusion coefficient with units $rad^2/s$.

Through the considerations of random forces acting on particles undergoing Brownian motion, Einstein and Smoluchowski independently arrived at the fluctuation-dissipation relation
\begin{equation} \label{flucdiss}
D_{(\theta)}=\frac{k_BT}{f_{(\theta)}},
\end{equation}
where $D_{(\theta)}$ can be either the translational or rotational diffusion coefficient, $k_B$ is the Boltzmann constant, $T$ is the temperature, and $f_{(\theta)}$ can be either the translational or rotational drag coefficient.\cite{Einstein,Smoluchowski} Under the simplifying assumption of spherical particles, Stokes related $f_{(\theta)}$ to the dynamic viscosity $\eta$ by
\begin{equation} \label{ftrans}
f=6 \pi \eta R,
\end{equation}
and by
\begin{equation} \label{frot}
f_{\theta}=8 \pi \eta R^{3},
\end{equation}
where $R$ is the probe particle's radius.\cite{Landau,stokes-einstein,detailed review of works} Combination of Eq.\eqref{ftrans} or \eqref{frot} with Eq.\eqref{flucdiss} directly relates $D$ or $D_{\theta}$ to $\eta$, yielding the well-known Stokes-Einstein relation.\cite{detailed review of works}

\subsection{\label{sec:level2}Scaling Law}

At lengthscales associated with individual random walkers, scaling laws for diffusion become suitable. Such is the case in crowded and disordered intracellular environments, where additional effective drag interactions between the probe and the individual solution molecules become important. Thus, the definition of an effective local viscosity, described by the diffusion coefficient and which accounts for these interactions, becomes relevant.\cite{microviscosity relevance} To introduce this effect and the introduction of active transport effects or Levy flights at the probe level and bridge the gap between the macroscopic and local viscosity scales, it is useful to consider the space-fractional version of the diffusion equation, Eq.\eqref{probdiff2}, given by
\begin{eqnarray}\label{fractional}
\frac{\partial P(\vec{r}, t)}{\partial t}=D \nabla^{n} P(\vec{r},t).
\end{eqnarray}
Here, $P(\vec{r},t)$ is the probability of finding a particle at a position $\vec{r}$ at time $t$, and $n \geqslant 0$ is a numerical factor which accounts for microscopic effects and alters the scaling of particle diffusion, where the fractional derivative is still a scalar operator.\cite{12 polymer physics 180-181} It is worth noting that in the limiting case of $n=2$ we retrieve the standard form of the diffusion equation. Again, the rotational version of Eq.\eqref{fractional} takes the same form. For intermediate cases, there is no known closed-form analytic solution. Instead, for diffusion in polymer solutions, one may use a stretched exponential model function to describe such intermediate cases. This function is found through considerations of the kinds of drag exerted by polymers on diffusing probes and on each other. Other variations of the diffusion equations are also use to describe the sub diffusive behavior with at time dependent form of the diffusivity, but this goes beyond of the current work. For our purposes, we are interest in describing scaling laws that pertain to the relationship of the probe and the solution, as explained further. 

Consider $f_{0}$, the drag experienced by a probe diffusing in an infinitely dilute polymer solution. As polymers are more rigid than the solvent, their presence in solution provides an additional drag term depending on their diffusion characteristics. Additionally, this increase is magnified at higher concentrations since the presence of polymers will effectively increase the drag experienced by other polymers as well. If we now consider two increments in polymer concentration, $dC$, then we can make the assumption that each increment increases $f$ by an amount $a(C)fdC$, where $a(C)$ is a concentration-dependent modifier to $f$. After the first increment, $f=f_{0}$ is then modified such that
\begin{equation}
f_{0} \rightarrow f_{0}+af_{0}dC=f(C).
\end{equation}
Then after the second increment, 
\begin{equation}
f_{0}+af_{0}dC \rightarrow (f_{0}+af_{0}dC)+a(f_{0}+af_{0}dC)dC=f(C+dC),
\end{equation}
or
\begin{equation}
f_{0}\rightarrow f_{0}(1+adC)\rightarrow f_{0}(1+adC)^{2}.
\end{equation}
Rearranging $f(C+dC)=f(C)(1+adC)$, and taking $df=f(C+dC)-f(C)$, we obtain
\begin{equation}
\frac{df}{dC}=af(C).
\end{equation}
Finally, taking the limit $dC\rightarrow 0$, this integrates to \begin{equation}
f(C)=f_{0}\exp(\int^{C}_{0}dCa(C)).
\end{equation}
Substituting this into Eq.\eqref{flucdiss} yields
\begin{equation} \label{almost}
D(C)=D_{0}\exp(-\int^{C}_{0}dCa(C)),
\end{equation}
where $D_{0}$ is the probe diffusion coefficient corresponding to $f_{0}$ at infinite polymer dilution. Assuming the hydrodynamic interactions between polymers to scale similarly to those for hard spheres described by Mazur and van Saarloos, the integral in Eq.\eqref{almost} can be shown to scale with $\beta C^{1-2x}$.\cite{mazur} Here, $\beta$ is an average of higher-order interactions determining how readily a solution's viscosity changes with polymer solution and $x$ is a scaling factor depending directly on the effective radii of gyration of both the polymers and the probe.\cite{derivation,mazur} Substituting $1-2x=\nu$, we finally obtain
\begin{equation} \label{eq:diffvis}
\frac{D}{D_{0}}=\exp\left({-\beta C^{\nu}}\right),
\end{equation}
the empirically verified universal scaling law for diffusion in polymer solutions.\cite{derivation,exponential, adler}
Such a scale-flexible relationship has been suggested through experiments under several polymer models, namely reptation-scaling treatment, hydrodynamic screening, and hydrodynamic scaling.\cite{reptation,hydrodynamic screening,hydrodynamic scaling} Combining with Eqs.\eqref{flucdiss}, \eqref{ftrans}, and \eqref{frot}, we obtain the normalized local viscosity as a function of osmolyte concentration, given by
\begin{equation} \label{eq:macromicrovis}
\frac{\eta}{\eta_{0}}=\exp \left({\beta C^{\nu}}\right),
\end{equation}
with $\eta_0$ as the viscosity at infinite dilution of the solute polymer. We use this key equation to determine the concentration dependence of viscosity, which is compared through various experimental methods that probe viscosity at various length- and time- scales.

\section{\label{sec:level1}Probing Viscosity Experimentally}

To experimentally verify the theoretical agreement between the macroscopic and the local viscosity, and to test the validity of Eq.\eqref{eq:diffvis}, we independently probe different timescales and lengthscales associated with each treatment. The macroscopic view of viscosity was assessed using rheometer measurements at a lengthscale of a few centimeters and a timescale close to seconds. Single particle tracking methods should be ideal to study viscosity from the microscopic perspective of Brownian motion; however, technical difficulties restrict most implementations of these methods for random walks in two dimensions.\cite{single part tracking} Therefore, alternative methods have been developed which simplify the determination of the diffusion coefficient. One of these methods is Fluorescence Correlation Spectroscopy (FCS), which probes the diffusion at sub-millisecond timescales and which can be implemented using inexpensive Field Programmable Gate Array (FPGA) boards as FCS hardware correlators. \cite{FCS method, FPGA} FCS is accomplished by correlating the time dependent fluorescence intensity as fluorescent markers diffuse through a small detection volume. Another method, time-resolved fluorescence anisotropy, allows for the study of rotational diffusion through the measurement of random re-orientations of the tracer particle. By considering the angular rotation of the particle, a similar viscous effect is observed. Thus, time-resolved anisotropy measures the tracer's rotational diffusion coefficient, allowing determination of viscosity at the nm lengthscale and ns timescale. \cite{anisotropy} Following is a brief introduction to these methods and brief descriptions of the materials used.

\subsection{\label{sec:level2}Glucose solutions and rhodamine-110 probe}

To probe the viscosity at different timescales and lengthscales, we created various solutions of D-Glucose (Table 1) ranging from $0\%$ to $30\%$ concentration in $5\%$ weight by volume increments in phosphate-buffered saline solution (PBS) containing 50mM Phosphate buffer and 150mM NaCl at pH 7.5. The concentration was then verified using a refractometer to have a specific index of refraction that corresponds to the expected Glucose concentration.

For steady state fluorescence spectroscopy, Rhodamine-110 (Table 1) was brought into solution and used at 2nM or 100nM solutions for FCS and time resolved measurements. Rhodamine-110 is a particularly bright fluorophore with a well-characterized fluorescence lifetime, making it an excellent candidate for these experiments. \cite{Rhodamine} 

\begin{figure}[h!]
\includegraphics[width=2.5in]{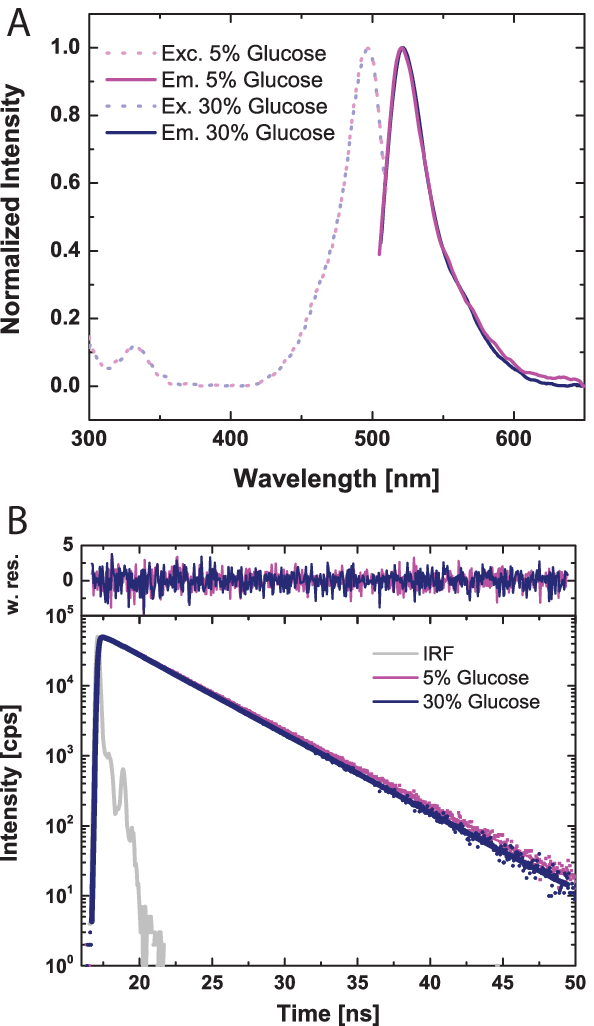}
\caption{(A) Fluorescence excitation and emission spectra of Rhodamine-110 in 5\% and 30\% (w/v) D-Glucose solutions. In both cases, the peak excitation wavelength was 487nm with peak emission wavelength at 521nm. (B) Corresponding time-resolved fluorescence decays and fit function Eq.\eqref{eq:timeresolveF}. Lifetimes for the 5\% and 30\% D-Glucose solutions exhibited the same behavior with lifetimes of 3.964 and 3.978 ns, respectively. Instrument response function (IRF) is shown in gray.}
\label{fig:spectra}
\end{figure}

For probe-based methods, the reporter molecules' fluorescent properties must not change under the conditions of the experiments. Thus, the steady-state fluorescence excitation and emission spectra at all concentrations of glucose were characterized.  The excitation and emission wavelengths for both the $5\%$ and $30\%$ D-Glucose solutions, seen in Fig. \ref{fig:spectra}A, were found to be 487 nm and 521 nm, respectively.

No major difference is observed in the normalized spectra, assuring that the fluorescence properties of Rhodamine-110 are independent of the environment. 

Spectral measurements are mostly insensitive to dynamic quenching, particularly if the concentrations of the solutions of Rhodamine-110 are not carefully controlled. Thus, to show that quenching does not occur or is minimally present in D-Glucose solutions, we measured the time-resolved fluorescence decays of Rhodamine-110 at all D-Glucose solution concentrations. Fig.(1B) shows representative normalized fluorescence decay spectra at 5\% and 30\% D-Glucose solutions, with he corresponding weighted residuals on top after the model function Eq.\eqref{eq:timeresolveF} is used for fitting.

The $5\%$ and $30\%$ solutions show very similar decays with no major changes in the fluorescence lifetimes derived from  Eq.\eqref{eq:timeresolveF}. Dynamic quenching would cause a shift towards shorter lifetimes as the concentration of the quencher increased.  This effect follows the Stern-Volmer relation.\cite{Stern-Volmer} From this, it was concluded that Rhodamine-110 suffered minimal collision-induced deactivation processes.

\subsection{\label{sec:level2}Rheometer}
\begin{figure}[h!]
\centering
\includegraphics[width=2.5in]{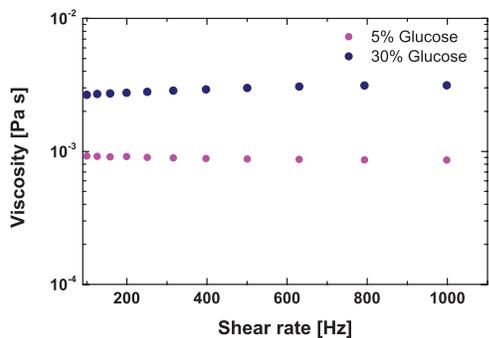}
\caption{Overlay of viscosity as a function of shear rate for Rhodamine 110 in 5\% and 30\%(w/v) glucose solutions. A torque range of $10^{-1}$ to $10^{2}$ $m$$N\cdot{m}$, a frequency range of $10^{-4}$ to $10^{2}$ $\textit{Hz}$, and an angular velocity range of $10^{-8}$ to $10^{2}$ $\frac{rad}{s}$ were used.}
\label{fig:rheology}
\end{figure}

Classical rheometry experiments consist of using a small amount of solution as a lubricant between two rotating plates. By measuring the resistance to flow imparted on the plates by the solution, the dynamic viscosity can be determined through Eq.\eqref{force}. 
Small amounts of D-Glucose solutions were placed between the two plates of a T.A. Instruments (Rheolyst model) AR1000-N Rheometer and measured using multiple shear rates in order to determine the viscosity. The typical response of a Newtonian fluid whose shear rate and shear stress follow a linear relationship is seen when we plot the viscosity in Fig. \ref{fig:rheology} as a function of the shear rate. This observed response gives a constant viscosity, indicating a lack of viscoelastic effects under varying levels of stress. As expected, the viscosity increased with the increase of D-Glucose concentration, and even at $30\%$ D-Glucose the solution still behaved as a Newtonian fluid. 

\subsection{\label{sec:level2}Fluorescence correlation spectroscopy}

Fluorescence correlation spectroscopy utilizes dilute solutions in the picomolar to micromolar range of diffusing particles to measure and correlate deviations from steady-state average fluorescence intensity in a small confocal volume. The diffusing particles either are fluorophores or are labeled with fluorophores, which are excited by lasers and subsequently re-emit light, giving rise to the intensity fluctuations. The correlation allows the determination of average particle numbers, photochemical effect timescales, diffusion times, and other parameters at the nanoscale. Choosing concentrations for measurement relies on a balancing of the factors that fluctuations from individual particles scale in the Poissonian distribution as $\frac{1}{\sqrt{\langle N \rangle}}$, where $\langle N \rangle$ is the time-averaged particle number in the confocal volume, and that the total fluorescence signal must be sufficiently high compared to noise signal.\cite{Schwille} Therefore, $\langle N \rangle$ on the order of a fraction of a particle to hundreds of particles is sufficient, depending on the used fluorophore.

The theory of FCS can be summarized as follows. Consider a total-time ($T$) averaged fluorescence signal 
\begin{equation}
\langle F(t) \rangle=\frac{1}{T}\int_{0}^{T}F(t)dt. 
\end{equation}
Then, deviations from the average fluorescence are given by
\begin{equation}
\delta F(t)=F(t)-\langle F(t) \rangle.
\end{equation}
This can be re-written as and integral over the effective detection volume
\begin{equation}
\delta F(t)=\kappa\int_{V}W(\mathbf{r})\delta(s(\mathbf{r}))dV,
\end{equation}
where $\mathbf{r}$ is the position vector, $W(\mathbf{r})$ is a function describing the geometry of the detection or confocal volume, and $\delta(s)$ characterizes individual fluctuation contributions due to parameters $s$ which can include fluctuations in quantum yield, absorption cross-sections, and, in our case, local concentration.\cite{Schwille} Plugging this into the autocorrelation function which describes the self-similarity of a signal after a time $\tau$, defined by
\begin{equation}
G(\tau)=\frac{\langle \delta F(t)\delta F(t+\tau)\rangle}{\langle F(t)\rangle^{2}},
\end{equation}
yields
\begin{equation}
G(\tau)=\frac{\int \int W(\mathbf{r})W(\mathbf{r}')\langle \delta(s(\mathbf{r}))\delta(s(\mathbf{r'}))\rangle dV dV'}{(\int W(\mathbf{r})\langle \delta(s(\mathbf{r})) \rangle dV)^{2}}.
\end{equation}
This equation can then be solved by quantifying the parameters $s$ for a given experimental setup, leading to the separability in $\delta(s)$ of contributing factors with sufficiently differing timescales and simplifying the problem.\cite{Schwille}

For these experiments we use a home built microscope adapted for confocal Fluorescence Correlation Spectroscopy measurements consists of a single diode laser (Model LDH-D-C- 485 at 485 nm; PicoQuant, Germany) operating at 40 MHz on an Olympus IX73 body. Freely diffusing Rhodamine-110 molecules are excited as they pass through the focal point of a 60X, 1.2 NA collar-corrected (0.17) Olympus objective. The power at the objective was set at 120 $\mu W$. The emitted fluorescence signal was collected through the same objective and spatially filtered using a 70 $\mu m$ pinhole to define an effective confocal detection volume. The emitted fluorescence was divided into parallel and perpendicular polarization components through band pass filters, ET525/50 for detecting “green” fluorescence photons. Two green PMA Hybrid detectors (PMA 40 PicoQuant, Germany) were used for photon counting. A time-correlated single photon counting (TCSPC) module (HydraHarp 400, PicoQuant, Germany) with a Time-Tagged Time-Resolved (TTTR) mode and two synchronized input channels were used for data registration and off-line fluorescence fluctuation analysis, similar to what has been used before.\cite{Dolino} Software correlation of TTTR data files was carried out to compute correlation curves that could cover over 12 orders of magnitude in time with a multi-tau algorithm.\cite{Felekyan}

It can be shown that the analytic solution to describe fluorescence correlation in our confocal setup is given as   
\begin{equation} \label{eq:FCS}
\begin{aligned}
G(t_c)=1+\frac{1}{N(1-x_T)}\cdot \frac{1}{1+\frac{t_c}{t_{diff}}}\cdot \frac{1}{\sqrt{1+\left( \frac{\omega_{xy}}{\omega_{z}}\right)^2\cdot\frac{t_c}{t_{diff}}}}\cdot \\ ( 1-x_T+x_T\exp( -t_c/t_T)),
\end{aligned}
\end{equation}
where $N$ is the mean number of molecules in the detection volume, $x_T$ is the fraction of molecules exerting triplet state kinetics with characteristic time $t_T$, $t_c$ is the correlation time, $t_{diff}=\omega_{xy}^2/(4D)$ is the diffusion time related to the geometrical parameter $\omega_{xy}$, which describes the detection volume along with $\omega_z$, $D$ is the diffusion coefficient, and 1 is used as the no-correlation baseline value.\cite{Elson-Magde} The detection volume is assumed to be a three-dimensional Gaussian in cartesian coordinates of the form
\begin{equation}
W(x,y,z) = \exp\left( -2(x^2+y^2)/\omega_{xy}^2\right)\exp\left( -2z^2/\omega_{z}^2\right),
\end{equation}
where the detection volume is defined by the $1/e^2$ radii denoted in terms of $\omega_{xy}$ and $\omega_z$. The ratio of $\frac{\omega_{xy}}{\omega_{z}}$ is used in calibration given a known standard. This Gaussian geometry is commonly used due to its ease of integration and accuracy in estimating the confocal volume.\cite{Elson-Magde} However, this is not necessary, and more complicated geometries can be introduced depending on the experimental setup, leading to a more complicated form of Eq.\eqref{eq:FCS}.\cite{ZMW-FCS}  

Both fluorescence triplet state kinetics and translational diffusion, resulting in local concentration fluctuations, were considered, allowing separability of $\delta(s)$ and leading to
\begin{equation}
G(\tau)=G_{diffusion}G_{triplet},
\end{equation}
where $G_{triplet}$ is the term in parentheses in Eq.\eqref{eq:FCS} multiplying the term corresponding to $G_{diffusion}$. Thus, Eq.\eqref{eq:FCS} represents the correlation function of a single tracer species diffusing in a chemically equilibrated solution detected in a confocal volume described by a three-dimensional Gaussian function.

Finally, the diffusion coefficient can be determined using the diffusion time $t_{diff}$ as
\begin{equation}
D=\omega_{xy}^2/(4t_{diff}).
\end{equation}
$D$ is again related to the viscosity through Eq.\eqref{flucdiss} and \eqref{ftrans}.

\begin{figure}[h!]
\centering
\includegraphics[width=2.5in]{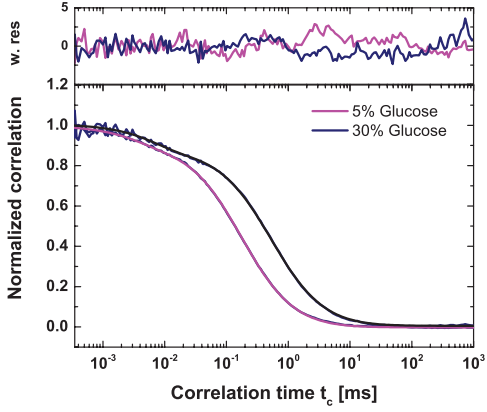}
\caption{Overlay of Fluorescence Correlation Spectroscopy curves and the corresponding fit model function Eq.\eqref{eq:FCS} for Rhodmaine-110 in 5\% and 30\% (w/v) D-Glucose solutions. The confocal volume's ratio of its height to its waist is given as a constant 4.609 for both concentrations. The time of diffusion for Rhodamine-110 was 0.192 ms for the 5\% solution and 0.597 ms for the 30\% solution. The fraction of triplet states, given by $x_{T}$, was 0.093 for 5\% and 0.130 for 30\%. The relaxation time $t_{T}$ for 5\% and 30\% are 0.003 to 0.008 ms, respectively.}
\label{fig:fcs}
\end{figure}

We use FCS, with Rhodamine-110 as our tracer particle, to determine the diffusion coefficient  as molecules travel across a confocal volume with a $1/e^2$ radius of $\sim$ 250 nm in sub millisecond timescales. As molecules traverse the confocal illumination volume, they emit light after being excited by a pulsed laser. When they exit the confocal volume the signal for that molecule stops. This causes fluctuations in intensity, which are recorded by the photon detectors. When the fluorescence intensity is correlated, these fluctuations generate a decay function that can be modeled with Eq. (\ref{eq:FCS}), from which the diffusion coefficient can be extracted. From this diffusion coefficient, the viscosity of the solution as sensed by the probe is found. Fig. \ref{fig:fcs} shows the correlation function and the model fits for Rhodamine-110 in $5\%$ and $30\%$ glucose solutions. The decrease in the diffusion constant, or increase in viscosity, at higher concentrations of glucose causes a shift in the characteristic correlation time towards longer correlation times. 

\subsection{\label{sec:level2}Time-resolved anisotropy}
The rotational diffusion is affected by the rotational drag force exerted on the probe by D-Glucose. This effect can be monitored by studying the rotational correlation time in time-resolved fluorescence anisotropy measurements. A brief theoretical treatment of this technique is as follows: consider a population of fluorophores, $n_{i}$, of species $i$. Then $n_{i,+}+n_{i,0}=n_{i}$, where $n_{i,+}$ is the number of particles in the excited state and $n_{i,0}$ is the number in the ground state. Given the fluorescence lifetime, $\tau_{i}$, and the excitation intensity, $I(t)$, the rate of change in the excited state population is given by the well-known rate equation,
\begin{equation}
\frac{dn_{i,+}}{dt}=-\frac{n_{i,+}}{\tau}+n_{i,0}I(t).
\end{equation}
Solving this equation for decay after an excitation pulse (such that $I(t>0)=0$ and we need only consider $t>0$) yields
\begin{equation}
n_{i,+}(t)=n_{i,+}(0)\exp(-\frac{t}{\tau_i}).
\end{equation}
Utilizing the facts that the fluorescence intensity is directly proportional to the number of excited states of fluorophores and that the total intensity is the sum of species contributions, the fluorescence intensity decay can be generally treated as a multi-exponential decay function using
\begin{equation} \label{eq:timeresolveF}
F(t)=\Sigma_{i} x_{i} \exp \left( {\frac{-t}{\tau_i}} \right).
\end{equation}
Here $x_{i}$ are the pre-exponential intensity factors.

For time-resolved anisotropy, the parallel and perpendicular decay components of the fluorescence, $F_{\parallel} (t)$ and $F_{\perp} (t)$, are considered by

\begin{eqnarray}
F_{\parallel}(t)= \frac{1}{3}F(t)\left[ 1+ 2r(t)\right],\\
F_{\perp}(t) = \frac{1}{3}F(t)\left[ 1- r(t)\right],
\end{eqnarray}
where $F(t)$ is the time-resolved fluorescence decay at magic angle conditions following Eq.\eqref{eq:timeresolveF} and $r(t)$ is the time-dependent anisotropy. In terms of the measured parallel and perpendicular components of the fluorescence decay, $r(t)$ is given by
\begin{equation}
r(t) = \frac{F_{VV}(t)-GF_{VH}(t)}{F_{VV}(t)+2GF_{VH}(t)},
\end{equation} 
\noindent where $G$ is an instrumental correction factor to account for changes in the the detection efficiency given wavelength and polarization. Here, $F_{VV}$ and $F_{VH}$ are $F_{\parallel}$ and $F_{\perp}$, respectively, after a correction for the differences in detection efficiency $G$ is made. \cite{Lakowicz} This G factor is easily measured using horizontally polarized excitation and the fluorescence decay in both polarization conditions.
Generally, it is possible to model the time-dependent anisotropy decay using a multi-exponential decay similar to Eq.\eqref{eq:timeresolveF}  given as
\begin{equation}\label{eq:aniso}
r(t)=\Sigma_{i} b_{i} \exp \left(\frac{-t}{\rho_i}\right),
\end{equation}
where $r_0 =\Sigma_{i} b_{i}$ is the fundamental anisotropy and $b_i$ are the fractional anisotropies that decay with correlation times $\rho_i$, which in turn are related to the rotational diffusion coefficient $D_{\theta}$. In the case of a sphere where there is only one rotational correlation time, $\rho$, this relation is given by
\begin{equation} \label{rottime}
\rho= \frac{1}{6D_{\theta}} =\frac{4 \pi \eta R^{3}}{3 k_{B} T},
\end{equation}
thus, relating the rotational correlation time directly to the dynamic viscosity as the particle rotates.
More complex expressions are predicted for nonsymmetric probe particles, but this goes beyond the scope of the exercise. In our case, we assume that the tracer particle, Rhodamine-110, is a sphere.

\begin{figure}[h!]
\centering
\includegraphics[width=2.5in]{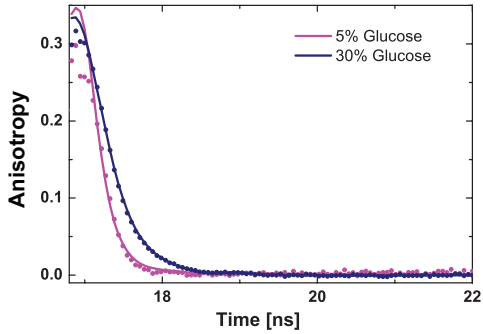}
\caption{Overlay of time-dependent anisotropy decays and the corresponding fit model function Eq.\eqref{eq:aniso} for Rhodmaine-110 in 5\% and 30\% (w/v) D-Glucose solutions. From this specific correlation, the rotational speeds in the 5\% and 30\% D-Glucose solutions were 100 $\pm 0.002$ ps and 385 $\pm 0.001$ ps, respectively.}
\label{fig:aniso}
\end{figure}

To assure that photophysical factors do not affect our probe experiments, we first measured the steady state fluorescence spectra, followed by ensemble time-correlated single-photon-counting (eTCSPC) using a Fluorolog3 spectrofluorometer in T-shape with a PDX detector (Horiba Yvon, USA) system. The light from a xenon lamp was used to collect excitation and emission spectra using an excitation monochromator set at 494 nm and an emission monochromator at 520 nm, accordingly. The scanning monochromator slit was set to 1 nm while the fixed monochromator slit was set to 5 nm. Spectra measurements were collected under magic angle conditions by using a vertical polarized excitation source and placing an emission polarizer to 54.7$^{\circ}$. 
For time-resolved fluorescence and time-resolved anisotropy measurements, the excitation source was a NanoLED 485 nm diode laser (NanoLED 485L, Horiba, Country) operating at 1 MHz. The emission monochromator slit was set to 16 nm (emission path). The signal of the photon counting unit was sent to a FluoroHub-B with a bin width of 55 ps. Fluorescence intensity decays in all polarizations were collected to determine the proper detection efficiency factor and determine time-resolved anisotropy. 
Fig. \ref{fig:aniso} shows the time-resolved anisotropy decays of Rhodamine-110 in $5\%$ and $30\%$ D-Glucose solutions. As expected, the rotational diffusion shows a slowing behavior in the anisotropy decay as the concentration of D-Glucose increases. By analyzing the decay function and fitting it with the model function, Eq.\eqref{eq:aniso}, the average rotational correlation time is found. Then, Eq.\eqref{rottime} is used to calculate the viscosity.

\subsection{\label{sec:level2}Error analysis}

Experimental reproducibility was evaluated using triplicated solutions of Rhodamine-110 in the FCS experiments. Each measurement was carried out at different times and with different starting stock conditions. The mean and standard deviation amongst these results was used to report our parameters. The uncertainty of experimental error was manifested using the standard deviation.

The statistical uncertainties of the fits from the time-resolved anisotropy decays were estimated by exploring the $\chi ^{2}$-surface of the model function given by Eq.\eqref{eq:chi2}. The error-margins of the individual fitting parameters are the projections from the individual parameter-dimensions. The maximum allowed $\chi ^{2}_{r, max}$ for a $1\sigma$ confidence interval is given by

\begin{equation} \label{eq:chi2}
\chi ^{2}_{r, max}=\chi ^{2}_{r, min} +  \sqrt[]{\frac{2}{N}}
\end{equation}

\noindent where $N$ is the number of fit points and $\chi ^{2}_{r, min}$ is the reduced chi-squared value of the best fit.\cite{confidence interval} All free fit parameters are varied simultaneously in a random manner. 

To evaluate the conditions of the experiments and how that introduces potential errors, we used the data collected with the Rheometer at different shear frequencies. Frequencies ranging from 0-1000Hz were used to evaluate the mean determined viscosity and standard deviation as representations of uncertainty in the methodology.

\section{\label{sec:level1}Discussion}

Different methodologies capture the drag frictional force at various time and length scales. For example, Fig. \ref{fig:eq7}A shows a linear correlation between the rotational and translational diffusion time. These measurements show a link between the interactions that occur at the nanoscale range to the interactions that occur over the span of hundreds of nanometers in the confocal illumination volume. This relationship between rotational and translational diffusion can further be seen through the fact that both measurements sense the same shear viscosity. Both methods do not exert significant stress on the fluid, thus it can be considered the ideal shear-less condition of a Newtonian fluid. While the rheomtery does induce shear stress, we noted that there is no dependence of the viscosity on this stress.  

\subsection{\label{sec:level2}Power law of viscosity}

\begin{figure}[h!]
\centering
\includegraphics[width=2.5in]{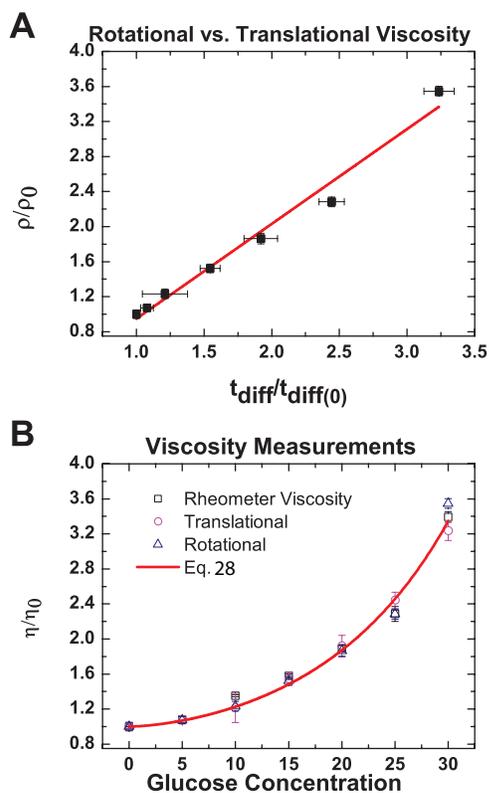}
\caption{Viscosity comparisons. A) A direct comparison of the measurement rotational motion with that of the translational motion of Rhodamine-110. B) A representation of the normalized viscosity change as a function of D-Glucose concentration for all three experiments.  The red curve represents the fit for averaged data, given by Eq.\eqref{eq:macromicrovis}.}
\label{fig:eq7}
\end{figure}

To reduce errors in calibration, the viscosity at different concentrations of Glucose was normalized to the viscosity of the buffer solution ($\eta/\eta_0$) (Fig. \ref{fig:eq7}B). Comparing the viscosity as derived from different experimental measurement, we observed that all values follow an exponential growth curve as a function of the concentration of D-Glucose. 
The exponential growth in Fig. \ref{fig:eq7}B was fit with Eq.\eqref{eq:macromicrovis} to evaluate the expected scaling law dependence of viscosity. Fitting the three data sets independently yields the results given in Table 2. Due to the apparent similarity in the data as shown in \ref{fig:eq7}A, the average of the fit parameters was calculated and the averaged data was also fit all together, yielding similar values also found in Table 2. Note that $\nu$ is specific to the dimensions and type of solute osmolyte and probe used, given the assumptions of Eq.\eqref{eq:macromicrovis} such as the dependence on the radii of gyration in $x$. However, we determined that this parameter is actually independent of dimensionality because the probe and probeless methodologies were able to show similar values for viscosity. The beta value is a measure of how readily the solution changes in viscosity based on its concentration.

\subsection{\label{sec:level2}Closing Remarks}

In summary, for a non-compressible Newtonian fluid, in which viscosity is independent of stress, probe viscosity (local microscopic) and shear viscosity (macroscopic) are identical. While probing the local viscosity, the probe does not exert stress while moving through the media; thus, no turbulent or convective effects must be accounted for, consistent with fluid behavior for low Reynolds number hydrodynamics. Combining classical rheology, FCS, and time-resolved anisotropy allows us to show that probe diffusion senses an effective local viscosity consistent with the macroscopic definition of shear viscosity. A similar approach can be applied to complex fluids, where the scaling behavior is relevant, and thus characterize the possible divergence between the local and macroscopic viscosities. Furthermore, these methodologies should be viable for future measurements of viscosities in compressible or other non-Newtonian fluids with either no or only small modifications to the experimental setup, assuming those materials are otherwise suitable for fluorescence techniques. For example, an external pressure source could be introduced to compress a sample in a T-shaped confocal setup for FCS and anisotropy measurements without interfering with the paths of the lasers to the sample or of light from the sample to the detectors. To this end, the modularity of such setups is a great aid. Additionally, classical rheometry experiments for comparison are also usable in pressure-sensitive experiments and, in fact, Maxwell's study of gas viscosities utilized such a setup with glass disks.\cite{Maxwell}

\section{\label{sec:level1}Acknowledgements}

We would like to thank Inna S. Yanez-Orozco for her technical assistance in the completion of this project. We also want to thank Dr. Foulger for facilitating us the use of the Rheometer. We acknowledge support from the Clemson University Creative Inquiry program.

\begin{center}
\begin{tabular}{|c|c|c|}
\hline
\multicolumn{3}{|c|}{Table 1. Compounds Used} \\
\hline
Structure& Name& Molecular Weight \\
\hline
\includegraphics[width=0.8in]{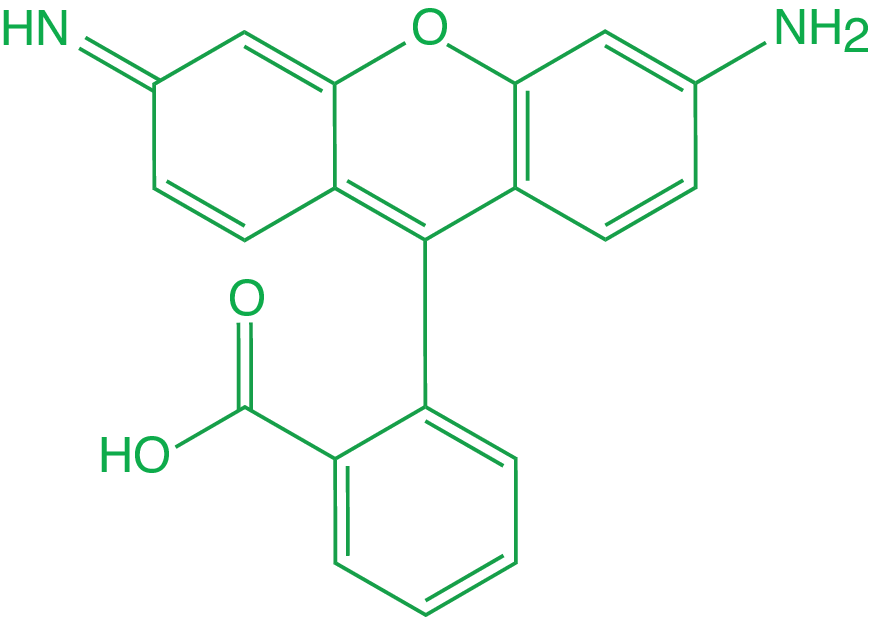} & Rhodamine-110& 366.80 \\
\hline
\includegraphics[width=0.8in]{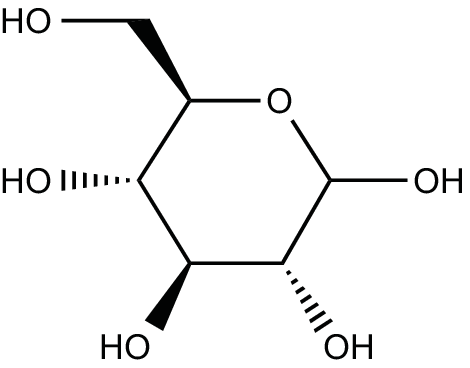} & D-Glucose& 180.16 \\
\hline
\end{tabular}\label{Table:materials}
\end{center}

\begin{center}
\begin{tabular}{|c|c|c|}
\hline
\multicolumn{3}{|c|}{Table 2. Fit Parameters} \\
\hline
Experiment& $\beta$ & $\nu$ \\
\hline
Rheometer&0.0133&1.32 \\
\hline
Translational&0.0079&1.47 \\
\hline
Rotational&0.0026&1.81 \\
\hhline {|=|=|=|}
Average&0.0079&1.53\\
\hline
Fit of Averaged Data&0.0049&1.62\\
\hline
\end{tabular}\label{Table:materials}
\end{center}

\end{document}